\documentclass[iop,ap]{emulateapj}

\usepackage{epsfig}
\usepackage[plainpages=false, colorlinks=true, anchorcolor=blue, citecolor=blue, linkcolor=blue, bookmarks=false]{hyperref}

\usepackage{graphics}
\usepackage{color}
\usepackage{endnotes}
\usepackage{natbib}
\usepackage{amssymb}
\usepackage{amsmath,bm}
\usepackage{times}
\usepackage{subfigure}
\usepackage{verbatim}

\newcommand{\beq}{\begin{equation}}
\newcommand{\eeq}{\end{equation}}
\newcommand{\bdm}{\begin{displaymath}}
\newcommand{\edm}{\end{displaymath}}
\newcommand{\bea}{\begin{eqnarray}}
\newcommand{\eea}{\end{eqnarray}}

\newcommand{\Msun}{M_\odot}
\def\lsim{\lower.5ex\hbox{$\; \buildrel < \over \sim \;$}}

\begin{document}

\title[FRBs and Radio Transients from Black Hole Batteries]{Fast Radio Bursts and Radio Transients from Black Hole Batteries}



\author{ Chiara~M.~F. Mingarelli}
\affiliation{TAPIR, MC 350-17, California Institute of Technology, Pasadena, California 91125, USA}
\affiliation{Max Planck Institute for Radio Astronomy, Auf dem H\"{u}gel 69, D-53121 Bonn, Germany }

\author{Janna Levin}
\affiliation{Institute for Strings, Cosmology and Astroparticle Physics (ISCAP), Columbia University, New York, NY 10027, USA}
\affiliation{Department of Physics and Astronomy, Barnard College, New York, NY 10027, USA}

\author{T.~Joseph~W. Lazio}
\affiliation{Jet Propulsion Laboratory, California Institute of Technology, Pasadena, California 91109, USA}

\begin{abstract}
Most black holes (BHs) will absorb a neutron star (NS) companion fully intact, without tidal disruption, suggesting the pair will remain dark to telescopes. 
Even without tidal disruption, electromagnetic luminosity is generated from the battery phase of the binary when the BH interacts with the NS magnetic field. Originally the luminosity was expected in high-energy X-rays or gamma-rays, however we conjecture that some of the battery power is emitted in the radio bandwidth.
While the luminosity and timescale are suggestive of fast radio bursts (FRBs; millisecond-scale radio transients) 
NS--BH coalescence rates are too low to make these a primary FRB source. Instead, we propose the transients form a FRB sub-population, distinguishable by a double peak with a precursor. The rapid ramp-up in luminosity manifests as a precursor to the burst which is $20\%-80\%$ as luminous, given 0.5~ms timing resolution. The main burst is from the peak luminosity before merger. The post-merger burst follows from the NS magnetic field migration to the BH, causing a shock. NS--BH pairs are especially desirable for ground-based gravitational wave (GW) observatories since the pair might not otherwise be detected, with electromagnetic counterparts greatly augmenting the scientific leverage beyond the GW signal. Valuably, the electromagnetic signal can break degeneracies in the parameters encoded in the GW as well as probe the NS magnetic field strength, yielding insights into open problems in NS magnetic field decay.
\end{abstract}

\keywords{black hole physics  --- gravitation --- gravitational waves ---  pulsars: general --- stars: neutron }
\maketitle

\section{Introduction}
Advanced ground-based interferometers are likely to detect gravitational waves (GWs) from compact binary coalescences within the next few years. Upgrades to aLIGO, e.g.~\cite{Harry:2010}, are complete and early observation runs have begun. In the coming years as aLIGO reaches peak sensitivity, they should be joined by VIRGO, and KAGRA to create a network of gravitational-wave observatories, e.g.~\cite{Virgo:2008, Somiya:2012}. The anticipation has motivated a closer look at the landscape of compact binary sources.
Encoded in the gravitational waveform is information about the source parameters, ranging from masses and spins to sky location. As such, a diverse and comprehensive toolkit is needed to extract the rich information available from these observations. While some GW source parameters are expected to be very well measured -- e.g., the binary chirp mass -- the source distance and inclination angle suffer from significant degeneracies, making them difficult to resolve, e.g.~\cite{LIGO:2013}. 

One of the most promising tools for lifting degeneracies in these parameters is the source's electromagnetic (EM) counterpart, e.g. \cite{nissanke+2013}. 
In a neutron star -- black hole (NS--BH) coalescence, if the NS is disrupted a relativistic jet powered by the rapid accretion of material onto the BH may produce a short gamma ray burst (GRB), e.g.~\cite{MetzgerBerger:2012, Nakar:2007}. GRBs are difficult to detect as the beamed emission results in a detection rate $<1$~yr$^{-1}$ with {\it Swift} for NS--BH and NS--NS mergers in the aLIGO/VIRGO volume. However, optical and radio afterglows can originate from the jet interaction with the medium surrounding the burst, lasting days to weeks, or weeks to months, respectively. 
Another EM counterpart is produced by the radioactive decay of heavy elements synthesized in the ejecta, powering an optical ``kilonova'', lasting a few days, see~\cite{LiPaczynski:1998}. 

However, most non-spinning BHs do not disrupt their companion NS before the plunge since the disruption radius, $r_\mathrm{tidal}=(M_\mathrm{BH}/M_\mathrm{NS})^{1/3}r_\mathrm{NS}=13~\Msun$ for $M_\mathrm{NS}=1.4~\Msun$ and $r_\mathrm{NS}=10$~km, is inside the Schwarzschild radius of a BH with $M_\mathrm{BH} \gtrsim 7\Msun$ (slightly larger for spinning BH). Hence the aforementioned EM follow-up techniques will not be achievable. Moreover, a GW detector with sensitivity comparable to the Einstein Telescope, e.g.~\cite{ET:2012} is needed to distinguish NS--BH and BH--BH binaries with the same mass ratio if the NS is not tidally disrupted, see~\cite{fbd13}. It is therefore advantageous to have an EM counterpart for a non-disrupted NS--BH binary, enabling the source identification of the GW signal and allowing for independent source parameter measurements, e.g.~\cite{tsang+2012}.

In this letter we describe a transient radio signal, typically lasting a few milliseconds with a luminosity of $\mathcal{L}\sim 10^{40}-10^{41}$~ergs/s, resulting from a BH interacting with a NS magnetic field in a configuration called a  ``battery'', see~\cite{glb69, Dong:2012, McWilliamsLevin:2011, Lyutikov:2011}, which may represent a fraction of the fast radio burst (FRB) population, e.g.~\cite{LorimerEtAl:2007}.

This transient has a distinctive signature: a rapid luminosity increase, manifesting as a precursor given sufficient signal-to-noise ratio and timing resolution, the ``burst'' from the peak luminosity before merger, and a post-merger burst at least 0.5~ms after the main burst, due to the migration of the NS magnetic field to the BH and subsequent magnetic field snapping.
For a sufficiently bright burst, the NS magnetic-field strength may be measured from the main and post-merger bursts, granting insights into the long-standing issue of NS magnetic-field decay.

\section{The Black Hole Battery}
\label{sec:BHBs}
In \cite{McWilliamsLevin:2011} a mechanism was proposed to light up a magnetized NS--BH binary for a few milliseconds when the BH moves through the NS dipole field. In this scenario, the BH acts like a battery, the NS acts like a resistor with its field lines as wires, and the charged particles in the NS magnetosphere are the current-carriers. The battery is established when the BH enters the closed magnetic dipole field lines within the light-cylinder of the spinning NS, defined as $2\pi r_\mathrm{L}/P=c$:  field lines inside the light cylinder are closed, those passing outside are open, e.g.~\cite{gj69}. For a spin period $P=1~s$, the battery connects when the pair are separated by $r_\mathrm{L}\sim 5\times 10^7$~m---thousands of Schwarzschild radii apart for a $10~M_\odot$ BH. However, the power may be unobservable until the final stages of coalescence, see Fig.~\ref{fig:rampup}. 

In the late inspiral regime, where GWs may be detectable with ground-based interferometers, 
the magnetic field threading the BH event horizon will be significant. 
In this paper we pursue the conjecture that the mechanism responsible for creating coherent broadband radiation in pulsars can similarly convert a fraction of the energy from the battery into the radio on a short timescale, similar to giant pulses from the Crab pulsar, e.g.~\cite{cordes+04}, creating rapid transient radio signals akin to FRBs, e.g.~\cite{Katz:2015}.
The radio efficiency parameter $\eta_r$ spans many orders of magnitude and increases with age. Recent radio pulsar observations from \citet{szm+14} show that the efficiency parameter $10^{-3}\lesssim \eta_r\lesssim 10^{-1}$ is appropriate for pulsars with $\tau\geq 10^7$~yrs. Guided by known radio pulsars and in the absence of other theoretical guidance, we adopt a fiducial value of $\eta_r=10^{-2}$.

 The luminosity of a BH battery is a function of the BH mass, $M_\mathrm{BH}$, the magnetic-field strength the NS at the poles, $B_p$, and the NS radius, $r_{NS}$, c.f.~\cite{McWilliamsLevin:2011}:
 
 \begin{eqnarray}
 \label{eq:generalBat}
 \mathcal{L}_{\mathrm{battery}} \!&=&  \!1.2\times 10^{41}~ \mathrm{erg/s}\left(\frac{\alpha v}{c}\right)^2 \left(\frac{B_p}{3\!\times\!10^{12}~\mathrm{G}}\right)^2 \!\!\left( \frac{\eta_{r}}{10^{-2}}\right)\! \!\! \nonumber \\
  &\times& \left(\frac{M_\mathrm{BH}}{10~\Msun}\right)^{2} \left(\frac{r_{NS}}{10~\mathrm{km}}\right)^{6} \left(\frac{r}{30~\Msun}\right)^{-6},
 \end{eqnarray}
where we have fixed the NS mass to be $1.4~\Msun$,  $\alpha\equiv\sqrt{1-2M_\mathrm{BH}/r}$ for a non-spinning BH, with $(\alpha v/c)^2\sim M/r$. The scaling above also depends on the unknown resistivities of the plasma and of the NS.
In Eq. \eqref{eq:generalBat}, $r(t)$ is the distance from the surface of the NS, and by substituting $(\alpha v/c)^2\sim M/r$, we implicitly assume a point-particle approximation. Since $\mathcal{L}_{\mathrm{battery}}\propto B^2$, and $B^2$ decays as $r^{-6}$, it is possible that when the BH gets very close to the NS surface that $\mathcal{L}_{\mathrm{battery}}$ could get a boost of orders of magnitude. Hence, Eq.~\eqref{eq:generalBat} is a conservative estimate of the power.
A spinning BH will also boost the luminosity, described in \cite{McWilliamsLevin:2011}.
 
NS magnetic-field decay is a long-standing issue in astrophysics. While e.g.~\citet{NarayanOstriker:1990, gob+02} showed that NS magnetic fields can decay on a timescale of $10^6$ -- $10^7$~yrs, statistical studies by e.g. \citet{Stollman:1987,lbh97, bwh+92} and simulations by \citet{fgk06} support decay time constants of $\geq 10^8$~yrs. Moreover, recent simulations by \cite{GourgouliatosCumming:2014} show that magnetic-field decay in middle-aged neutron stars is dramatically slowed, and may therefore differ from what is currently drawn on $P$--$\dot P$ diagrams. We therefore take as plausible that NS magnetic fields do not significantly decay and consider $B_p \sim 3\times10^{12}\,\mathrm{G}$ as a fiducial value.
In fact, since $\mathcal{L}_\mathrm{battery}\propto B^2_p$, {\it cf.} Eq. \eqref{eq:generalBat}, electromagnetic observations of BH batteries may provide an avenue to probe the magnetic-field strength of old NS.

\section{Gravitational Waves from NS--BH binaries}
\label{sec:gws}

Compact binary coalescences are the most promising sources of GWs for ground-based interferometers, e.g.~\cite{Harry:2010, Virgo:2008, Somiya:2012}. 
 These detectors operate in the high frequency GW regime, with peak sensitivity between 50 and 1000~Hz. 
Let us consider a NS--BH binary with $M_\mathrm{BH}=10~\Msun$ and $M_\mathrm{NS}=1.4~\Msun$. Such a binary has a chirp mass $\mathcal{M}_c=\mu^{3/5}M^{2/5}=3~\Msun$, where $\mu=m_1m_2/M$ is the reduced mass of the binary and $M=m_1+m_2$ is its total mass. 
We are interested in the last few milliseconds before coalescence, for reasons described in Sec. \ref{sec:frb}, therefore the binary's GW frequency and time to coalescence are reported here at the innermost stable circular orbit $6~M_\mathrm{BH}$.
The binary separation $r(t)$ is computed via~\citet{Peters1964}: 
\begin{equation}
\label{eq:sep}
r(t)=\left(\frac{256}{5}\mu M^2 \right)^{1/4} (t_c-t)^{1/4}\, 
 \end{equation}
hence, for a binary $M=11.4~\Msun$ separated by $r(t)=6~M_\mathrm{BH}$, the GW frequency is 470~Hz. The time to coalescence $t_c$ from $6~M_\mathrm{BH}$ is obtained via Eq. \eqref{eq:sep}:
\beq
\label{eq:tc}
t_c = 7.8~ \mathrm{ms}\left( \frac{r(t)}{6~M_\mathrm{BH}}\right)^{4} \left( \frac{\mu}{1.2~\Msun }\right)^{-1}\left( \frac{M}{11.4~\Msun}\right)^{-2} \, .
\eeq
The detection rates for ground-based interferometers depend on the expected rates of compact binary coalescence events. 
Due to the lack of direct EM observations of compact binary systems containing BHs, NS--BH rates are based on population-synthesis models, e.g.~\cite{LIGOparamReview:2010}. 
The recent local merger rate for NS--BH binaries with $\mathcal{M}_c=3.2~\Msun$ is $3$ -- $20$~Gpc$^{-3}$ yr$^{-1}$, see~\cite{DominikEtAl:2014}, with the expected detection rate of 1 -- 6~yr$^{-1}$ with aLIGO and 2 -- 15~yr$^{-1}$, using a 3-detector network. For $\mathcal{M}_c=3\Msun$, the detection probability scales by 0.75 for by aLIGO, see Fig. 6 of \cite{DominikEtAl:2014}. 

\begin{table*}
\centering
\begin{tabular}{lclllllll}
\hline
         & Distance (Gpc) & $S_\nu$ (Jy) &$\mathcal{L}_\mathrm{radio}$ (erg/s) & Ref \\
          \hline
 FRB 010724    & 1.0         & 30          & $5.03 \times 10^{43}$ &  \cite{LorimerEtAl:2007}\\    
 FRB  110220   &   2.8     &  1.3        &    $1.71\times 10^{43}$                 &\cite{ThorntonEtAl:2013}\\
 FRB  110703   &    3.2     & 0.5        &         $8.58\times10^{42}$                &\cite{ThorntonEtAl:2013}\\
 FRB 131104    &   1.0      & 2.0  & $3.35\times 10^{42}$    & \cite{RaviEtAl:2014} \\
FRB  110627   &   2.2      & 0.4        &       $3.24\times 10^{42}$                   &\cite{ThorntonEtAl:2013}\\
FRB  120127   &    1.7     &  0.5       &       $2.42\times10^{42}$                   &\cite{ThorntonEtAl:2013}\\
 FRB 140514 	& 1.7		&0.47	& $2.28\times 10^{42}$			& \cite{PetroffEtAl:2015} \\
 FRB 011025   & 2.1         & 0.3         &  $2.22\times 10^{42}$			& \cite{BurkeSpolaorBannister:2014} \\
 FRB 121102 	& 1.0		&0.4		&$6.70\times 10^{41}$ 			&\cite{SpitlerEtAl:2014} \\
FRB 010621		& 0.7	&0.4		& $3.28\times 10^{41}$ 						&\cite{KeaneEtAl:2012}
\end{tabular}
\caption{Luminosities of reported fast radio bursts, {\it cf.} Eq. \eqref{eq:FRBlum}, are comparable to Eq. \eqref{eq:generalBat}. The current population does not, however, show the distinctive precursor or double peak of the BH battery. So far, all FRBs have been found at 1.4~GHz with distances inferred from their dispersion measure-induced frequency sweeps. The luminosity from \cite{RaviEtAl:2014} is drawn from their Fig. 3 at 1432~MHz, and for the \citet{KeaneEtAl:2012} burst, we fix $h=0.7$ for illustrative purposes.}
\label{Table1}
\end{table*}

Host identification and parameter estimation are difficult tasks, e.g.~\cite{vrf+15}, especially for non-disrupted and non-spinning NS--BH mergers. Numerically, the differences between a BH--BH and a NS--BH gravitational waveform, orbital evolution, and characteristics of the final remnant cannot be resolved if they have the same mass ratio, see~\cite{fbd13}. In fact, \citet{fbd13} claim that only an EM counterpart could prove the presence of a NS in low-spin systems, until the advent of GW detectors with a sensitivity comparable to the proposed Einstein Telescope, e.g.~\cite{ET:2012}. 

\section{Fast Radio Bursts from BH Batteries}
\label{sec:frb}

\begin{figure*}[ht]
     \begin{center}

        \subfigure[ \, BH Battery luminosity, 1ms resolution] {%
		\label{fig:lum1ms}
		\includegraphics[scale=1]{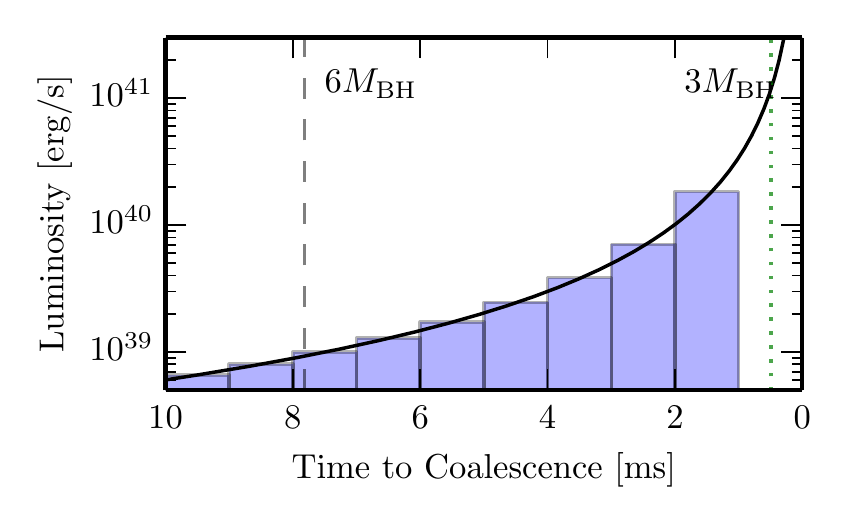}
        }
	\subfigure[ \,  BH Battery luminosity, 0.5 ms resolution]
	{
		\label{fig:lum5ms}
	\includegraphics[scale=1]{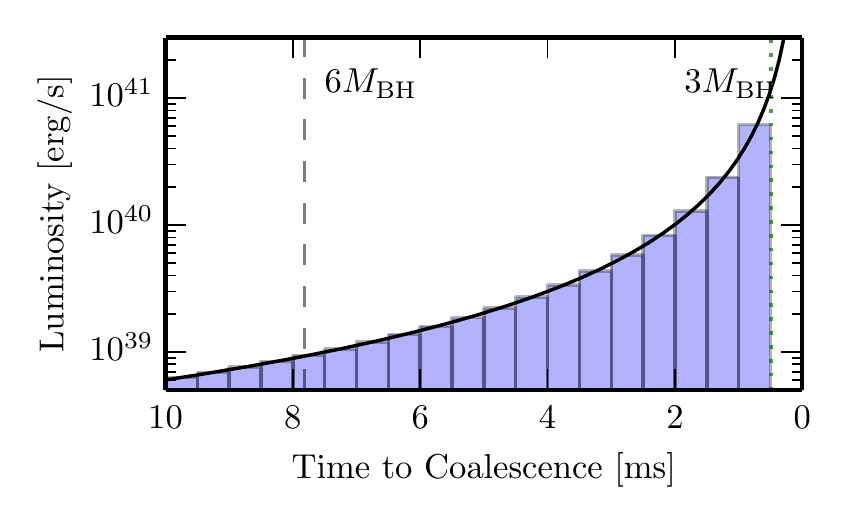}
	}
	\subfigure[ \, BH Battery luminosity for $M_\mathrm{BH}=10~\Msun$, $B=10^{15}$~G]
	{
		\label{fig:magnetar}
	\includegraphics[scale=1]{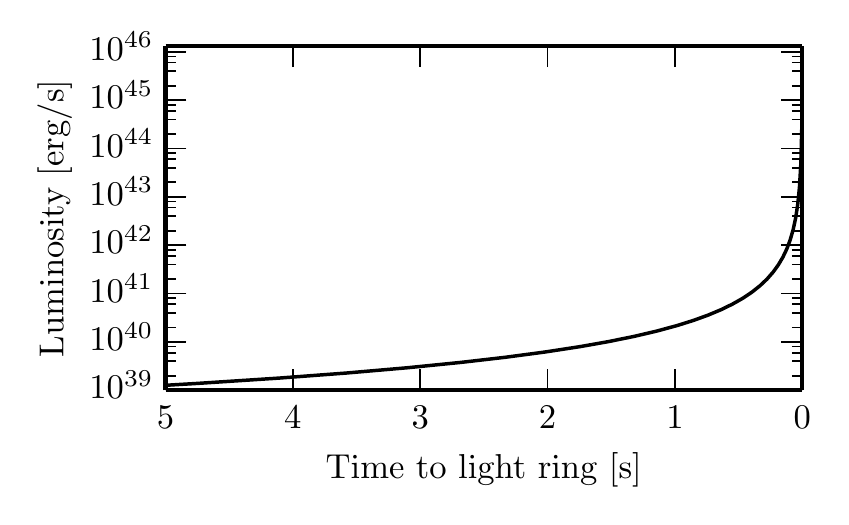}
	}
	\subfigure[ \, Relative luminosity of precursor to main burst at light ring]
	{
		\label{fig:LR_ratio}
	\includegraphics[scale=1]{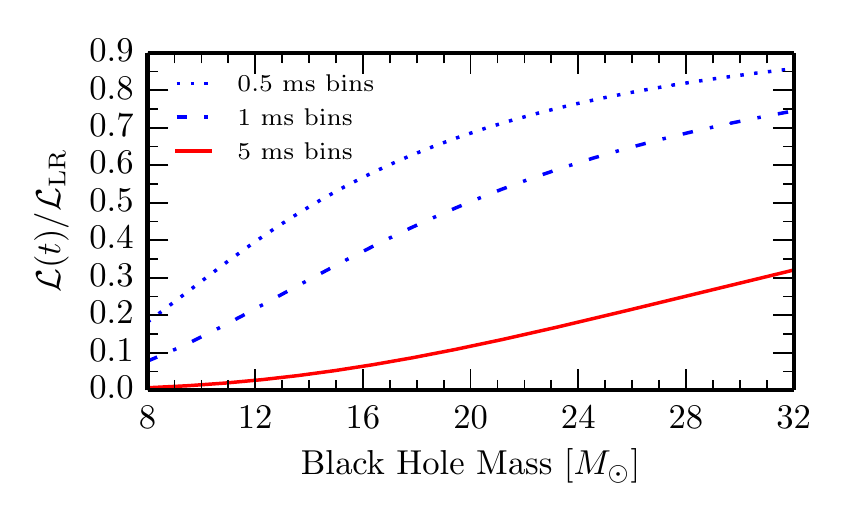}
	}
\end{center}

	\caption{Luminosity of NS--BH binary. In panels (a) and (b) $M_\mathrm{BH}=10~\Msun$ and $B=3\times 10^{12}$~G. Bars are the integrated luminosity in each bin. Rapid increase in luminosity milliseconds before coalescence may generate a precursor, given sufficient bin resolution. In panel (a) we do not include the final millisecond, since the energetics are highly uncertain, and in (b) we halt at $3~M_\mathrm{BH}$. For a magnetar with $B=10^{15}$~G, the BH battery luminosity would be visible for over 5~s, panel (c). While the luminosity depends on all the parameters in Eq. \eqref{eq:generalBat}, the relative strength of the precursor to the burst luminosity is only a function of the BH and NS masses, reported at a separation $r(t)$. In panel (d) we fix $r(t)=3~M_\mathrm{BH}$ and $M_\mathrm{NS}=1.4~\Msun$ and show that for 0.5~ms bins, one can detect precursors with luminosities $\sim 20\%-80\%$ of the peak (burst) luminosity. The noise in the telescope is proportional to $1/\sqrt{\Delta t}$,~\cite{LorimerKramer:2012}, where $\Delta t$ is the time step (bin size), hence one gains $\sqrt{2}$ in sensitivity moving from 0.5~ms to 1~ms bins, at the expense of the visibility of the precursor.	}
\label{fig:rampup}	 
\end{figure*}

Fast radio bursts (FRBs) are a recently identified class of single, coherent, millisecond radio pulses, e.g.~\cite{LorimerEtAl:2007, KeaneEtAl:2012,  ThorntonEtAl:2013, RaviEtAl:2014, BurkeSpolaorBannister:2014, SpitlerEtAl:2014, PetroffEtAl:2015, katz:2014}.  They are marked by the characteristic frequency sweep indicative of propagation through a cold plasma, but with a dispersion measure (DM) or total electron column density, suggesting an origin from extragalactic compact objects, e.g. \cite{CordesLazio:2002, LorimerEtAl:2007, KeaneEtAl:2012, ThorntonEtAl:2013, LuanGoldreich:2014, RaviEtAl:2014}. Several candidates have been proposed as FRB sources, including NS mergers, e.g. \cite{LiPaczynski:1998, HansenLyutikov:2001, Totani:2013}, ``supramassive'' NS collapse \cite{FalckeRezzolla:2014} and giant pulses or bursts from magnetars, e.g. \cite{PopovPostnov:2013, pp07}. 

So far nine FRBs have been found at Parkes by \cite{LorimerEtAl:2007, ThorntonEtAl:2013, BurkeSpolaorBannister:2014, RaviEtAl:2014, PetroffEtAl:2015} and one was found at the Arecibo Radio Telescope by \cite{SpitlerEtAl:2014}. To determine if the hypothesized short-lived burst of radio waves from the battery phase of NS--BH binaries may form a sub-population of FRBs, we examine the energetics of known FRBs in Table \ref{Table1} and compare to Eq. \eqref{eq:generalBat}. 

We take the characteristic flux density of a FRB to be $S_\nu = 1$~Jy and assume that it is at a distance $D = 1$~Gpc, see Table \ref{Table1}.  There is essentially no spectral information currently available on FRBs, with all of the published FRBs having been detected at~1.4~GHz.  We therefore assume that they are relatively broadband emitters, similar in nature to radio pulsars, with a typical bandwidth~$\Delta\nu$ comparable to the frequency of emission (i.e. $\Delta\nu/\nu \sim 1$).  Finally, with little information available on possible beaming angles, we assume $\Omega \sim 1$~sr.  We find that the typical luminosity is then 

\begin{equation}
\label{eq:FRBlum}
\mathcal{L}_\mathrm\!=\!1.3 \times 10^{41} \mathrm{erg/s}\left(\frac{S_\nu}{1~\mathrm{Jy}}\right)\!\!\left(\frac{\Delta\nu}{1.4~\mathrm{GHz}}\right) \!\! \left(\frac{\Omega}{1~\mathrm{sr}}\right) \!\!\left(\frac{D}{1\,\mathrm{Gpc}}\right)^2 \!\!\!.
\end{equation}

While FRBs have the required luminosity to have originated from BH batteries, the rates for FRBs -- roughly $2.3\times10^4$~yr$^{-1}$ Gpc$^{-3}$, e.g.~\cite{Totani:2013} -- are three orders of magnitude larger than the optimistic estimate of NS--BH merger rate of 20 yr$^{-1}$ Gpc$^{-3}$, see~\cite{DominikEtAl:2014}. 
Our claim is not that all FRBs originate from BH batteries, rather that there could be multiple populations of FRBs (analogous to the multiple populations---long and short---for GRBs), with some of the FRBs being the radio counterpart to a non-disrupted NS--BH binary coalescence, possibly associated with a GW event. 

While these sources may be rare, the Canadian Hydrogen Intensity Mapping Experiment (CHIME) is being outfitted with a fast-detection backend with projected rates approaching $10^4$~yr${}^{-1}$ (V.~Kaspi~2015, private comm.), and a proposed augmentation of the Very Large Array's processing capability would enable discovery rates approaching $10^3$~yr${}^{-1}$ (C.~Law~2015, private comm.).  A modest estimate is that as many as 5000 FRBs could be discovered over the next 5~years.

The fast radio transients we describe are distinctive. Firstly, there is a ramp-up in the luminosity due to the NS orbiting and plunging into the BH. The continuous luminosity increase may appear as at least one precursor when binned, Fig. \ref{fig:rampup}. 
We estimate the luminosity until the light ring at $3~M_\mathrm{BH}$, since the applicability of Eq. \eqref{eq:FRBlum} breaks down at close separations. In fact, though winding still, the NS begins to plunge between $6~M_\mathrm{BH}$ and $3~M_\mathrm{BH}$. However, in principle the power only surges as the NS approaches the horizon. In fact, \cite{McWilliamsLevin:2011} show that systems with significant BH spin are even more luminous than non-spinning ones.

The relative strength of the precursor to the burst is easily computed: $\mathcal{L}(t)/\mathcal{L}_0 = [r(t)/r_0]^{-7}$, and substituting Eq. \eqref{eq:sep} gives $\mathcal{L}(t)/\mathcal{L}_0 = (\Delta t/\Delta t_0)^{-7/4}$, where $\Delta t= t_c-t$. For example, we compare the luminosity of a canonical $10~\Msun$ BH and 1.4~$\Msun$~NS BH battery at the light ring to the luminosity 0.5~ms earlier. Using Eq. \eqref{eq:tc}, $t_c\simeq0.5$~ms and $\mathcal{L}(t)/\mathcal{L}_0  =0.29$. The luminosity of the precursor is therefore $\sim 30\%$ of the burst. If such a burst were detected with S/N of 30, the precursor would thus have a S/N $=9$. In Fig. \ref{fig:LR_ratio}, the relative signal strength is so computed for a range of BH masses. We find that the precursor can be between $20\%-80\%$ of the luminosity of the main burst, depending on bin resolution.
The current FRB population has been detected with $\mathrm{S/N} \gtrsim 10$, e.g.~\cite{LorimerEtAl:2007, BurkeSpolaorBannister:2014}.  While none show evidence of a precursor, even higher S/N is likely required to detect the precursor. 

The second signature of this FRB sub-population is a post-merger burst from the migration of the NS magnetic field to the BH and the subsequent violent magnetic-field snapping. This second peak is also predicted by the magnetic shock from a supramassive NS collapse into a BH, called a ``blitzar'' from~\cite{FalckeRezzolla:2014}. In both the battery and blitzar model, a shock travels outwards and produces radio emission which is in turn modulated by the ringdown of the BH event horizon, resulting in exponentially decaying sub-ms radio pulses. For the battery model, the delay between the main FRB and the post-merger burst is at least the sum of the time to coalescence and the light-crossing time of the BH, e.g for $M_\mathrm{BH}=10~\Msun$ is $\sim 500~\mu$s + 50~$\mu$s $> 0.5$~ms. It is probably much longer since the magnetized BH could retain a magnetosphere which itself supports the magnetic field for longer, e.g.~\cite{Lyutikov:2011}. This admittedly naive estimate reinforces the importance of 0.5~ms resolution, so that one can in principle resolve even this (these) post-merger burst(s) from magnetic field snapping and BH ringdown. 

The luminosity of the post-merger peak depends on the NS period $t_{NS}$, magnetic field strength and radius, as well as the fraction of magnetic field energy available for the burst, $\eta_B$, see Eq. (4) of~\citet{FalckeRezzolla:2014}, and is potentially as bright as the main burst:

\begin{eqnarray}
\label{eq:post-burst}
\mathcal{L}_{\mathrm{post}} \simeq 3.8&\times& 10^{41} \mathrm{erg/s} \left( \frac{\eta_\mathrm{B}}{0.05} \right) \left( \frac{B_p}{3\times 10^{12}~\mathrm{G}} \right)^2 \nonumber \\
&\times& \left( \frac{r_{NS}}{10~\mathrm{km}} \right)^3 \!\! \left( \frac{t_{NS}}{1~\mathrm{s}} \right)^{-1}.
\end{eqnarray}

\section{Discussion}
\label{sec:discussion}
The population of FRBs described here may have a double peak as well as a precursor: the precursor is from the binned ramp-up in luminosity, followed by the main burst at maximum luminosity, and a post-merger burst due to magnetic field shock. 
Figure \ref{fig:rampup} shows examples of the luminosity of the BH battery milliseconds before coalescence. When observed with sufficient time resolution, $\sim0.5$~ms, and signal-to-noise, the notable precursor feature emerges. 
The post-merger burst can manifest due to the migration of the NS magnetic field to the BH at the time of coalescence, and the subsequent magnetic field shock, similar to a FRB from the ``blitzar'' model. This post-merger burst would occur at least $\sim0.5$~ms after the main FRB, for a $10~\Msun$ BH, and depending on the NS's intrinsic parameters, and may be as luminous as the main burst, see Eq. \eqref{eq:post-burst}. While we consider a non-spinning BH in this study, spin would make the system even more luminous and change the innermost stable circular orbit from $6~M_\mathrm{BH}$ to $M_\mathrm{BH}$, if prograde.
The luminosity ramp-up itself may contain some structure pulsed at the NS orbital frequency. This would be most interesting immediately before merger, thus at the millisecond level, and would therefore require a very strong EM signal to be detectible.

The distance to a FRB is estimated using the DM, and are in fact upper bounds with an uncertainty of around $20\%$, e.g.~\citet{ThorntonEtAl:2013}. If the sources are closer, the radio emission efficiency and other parameters in Eqs.~\eqref{eq:generalBat} and \eqref{eq:FRBlum} could be significantly smaller. 

The most recent merger rates for NS--BH systems with $\mathcal{M}_c=3.2~\Msun$ range from 3 -- 20~Gpc$^{-3}$ yr$^{-1}$, of which 2 -- 15~yr$^{-1}$ are expected to be detected via GW emission with a 3-detector network, see~\cite{DominikEtAl:2014}. If $\mathcal{M}_c=3~\Msun$, $75\%$ of these sources are expected to be detected. Some of these may have EM counterparts from the NS disruption, while the majority do not and may belong to the NS--BH population described here.  
An EM counterpart from a NS--BH coalescence allows for an independent BH and NS mass measurement, Eq. \eqref{eq:generalBat}, as well for a complementary distance measurement to the source, if other parameters are sufficiently well constrained. 
In the event of a NS--BH GW detection, a coincident EM detection may be possible via new low-latency pipelines sending out triggers, outlined in e.g. \cite{nissanke+2013, chu+2015}, or if the telescope is already pointed at the source. This is most probable for telescopes such as CHIME, which shares $\sim20\%$ of the two-detector LIGO network sky localization arc from~\cite{kasliwalNissanke:2014}, Fig. 2. While we rely on sky localization arcs from \cite{kasliwalNissanke:2014}, who consider NS--NS binaries, a comparable S/N for a NS--BH binary would have similar sky localization prospects. In fact, coincident detection prospects with low-frequency radio arrays are especially tantalizing, as dispersion may delay the radio signal, further enabling a possible LIGO/Virgo trigger; \cite{yba+15, trott:2013}. 

In addition to the coherent radio emission, we also expect a burst of synchrocurvature radiation in X-rays and/or gamma-rays, which may be more difficult to detect. However, if the power is heavily reprocessed with longer radio emission timescales---perhaps even much after an initial x-ray or gamma-ray burst---then we are looking for a new kind of radio transient. Indeed, the brief transient EM radiation from the BH battery may be detectible in many wavelengths, and as such may be detected first and more frequently than the GW signal, thereby informing NS--BH merger rates. 

Even without a GW counterpart, the radio transients from NS--BH binaries offer a unique avenue to explore the properties of the cosmic population of NS--BH binaries and potentially measure the magnetic field of old NS. In the age of transient searches, we encourage observers to consider these fascinating source candidates. 

\section*{acknowledgments}
\begin{acknowledgments}
We thank the referee, Michele Vallisneri and Samaya Nissanke for carefully reading the manuscript. We acknowledge valuable discussions with Sean McWilliams, Mansi Kasliwal, David Tsang, Andrea Lommen, Stephen Taylor, Justin Ellis, Jocelyn Bell-Burnell, Peter Goldreich and Francesco Pannarale. CMFM was supported by a Marie Curie International Outgoing Fellowship within the European Union Seventh Framework Programme. JL thanks the Tow Foundation for their support. JL was also supported by a Guggenheim Fellowship. Part of this research was carried out at the Jet Propulsion Laboratory, California Institute of Technology, under a contract with the National Aeronautics and Space Administration. An ipython notebook which reproduces our results is available \url{https://github.com/ChiaraMingarelli}.
\end{acknowledgments}

\bibliographystyle{apj}


\begin{thebibliography}{}
\expandafter\ifx\csname natexlab\endcsname\relax\def\natexlab#1{#1}\fi

\bibitem[{Aasi {et~al.}(2013)Aasi, Abadie, Abbott, Abbott, Abbott, Abernathy,
  Accadia, Acernese, Adams, Adams, Addesso, Adhikari, Affeldt, Agathos,
  Agatsuma, Ajith, Allen, Allocca, Amador~Ceron, Amariutei, Anderson, Anderson,
  Arai, Araya, Ast, Aston, Astone, Atkinson, Aufmuth, Aulbert, Aylott, Babak,
  Baker, Ballardin, Ballmer, Bao, Barayoga, Barker, Barone, Barr, Barsotti,
  Barsuglia, Barton, Bartos, Bassiri, Bastarrika, Basti, Batch, Bauchrowitz,
  Bauer, Bebronne, Beck, Behnke, Bejger, Beker, Bell, Bell, Belopolski,
  Benacquista, Berliner, Bertolini, Betzwieser, Beveridge, Beyersdorf,
  Bhadbade, Bilenko, Billingsley, Birch, Biswas, Bitossi, Bizouard, Black,
  Blackburn, Blackburn, Blair, Bland, Blom, Bock, Bodiya, Bogan, Bond,
  Bondarescu, Bondu, Bonelli, Bonnand, Bork, Born, Boschi, Bose, Bosi, Bouhou,
  Braccini, Bradaschia, Brady, Braginsky, Branchesi, Brau, Breyer, Briant,
  Bridges, Brillet, Brinkmann, Brisson, Britzger, Brooks, Brown, Bulik, Bulten,
  Buonanno, Burguet\char21{}Castell, Buskulic, Buy, Byer, Cadonati, Cagnoli,
  Calloni, Camp, Campsie, Cannon, Canuel, Cao, Capano, Carbognani, Carbone,
  Caride, Caudill, Cavagli\`a, Cavalier, Cavalieri, Cella, Cepeda, Cesarini,
  Chalermsongsak, Charlton, Chassande-Mottin, Chen, Chen, Chen, Chincarini,
  Chiummo, Cho, Chow, Christensen, Chua, Chung, Chung, Ciani, Clara, Clark,
  Clark, Clayton, Cleva, Coccia, Cohadon, Colacino, Colla, Colombini, Conte,
  Conte, Cook, Corbitt, Cordier, Cornish, Corsi, Costa, Coughlin, Coulon,
  Couvares, Coward, Cowart, Coyne, Creighton, Creighton, Cruise, Cumming,
  Cunningham, Cuoco, Cutler, Dahl, Damjanic, Danilishin, D'Antonio, Danzmann,
  Dattilo, Daudert, Daveloza, Davier, Daw, Dayanga, De~Rosa, DeBra, Debreczeni,
  Degallaix, Del~Pozzo, Dent, Dergachev, DeRosa, Dhurandhar, Di~Fiore,
  Di~Lieto, Di~Palma, Di~Paolo~Emilio, Di~Virgilio, D\'{i}az, Dietz, Donovan,
  Dooley, Doravari, Dorsher, Drago, Drever, Driggers, Du, Dumas, Dwyer, Eberle,
  Edgar, Edwards, Effler, Ehrens, Endr\ifmmode~\mbox{\H{o}}\else \H{o}\fi{}czi,
  Engel, Etzel, Evans, Evans, Evans, Factourovich, Fafone, Fairhurst, Farr,
  Farr, Favata, Fazi, Fehrmann, Feldbaum, Feroz, Ferrante, Ferrini, Fidecaro,
  Finn, Fiori, Fisher, Flaminio, Foley, Forsi, Forte, Fotopoulos, Fournier,
  Franc, Franco, Frasca, Frasconi, Frede, Frei, Frei, Freise, Frey, Fricke,
  Friedrich, Fritschel, Frolov, Fujimoto, Fulda, Fyffe, Gair, Galimberti,
  Gammaitoni, Garcia, Garufi, G\'asp\'ar, Gelencser, Gemme, Genin, Gennai,
  Gergely, Ghosh, Giaime, Giampanis, Giardina, Giazotto, Gil-Casanova, Gill,
  Gleason, Goetz, Gonz\'alez, Gorodetsky, Go\ss{}ler, Gouaty, Graef, Graff,
  Granata, Grant, Gray, Greenhalgh, Gretarsson, Griffo, Grote, Grover,
  Grunewald, Guidi, Guido, Gupta, Gustafson, Gustafson, Hallam, Hammer,
  Hammond, Hanks, Hanna, Hanson, Harms, Harry, Harry, Harstad, Hartman, Haster,
  Haughian, Hayama, Hayau, Heefner, Heidmann, Heintze, Heitmann, Hello,
  Hemming, Hendry, Heng, Heptonstall, Herrera, Heurs, Hewitson, Hild, Hoak,
  Hodge, Holt, Holtrop, Hong, Hooper, Hough, Howell, Hughey, Husa, Huttner,
  Huynh-Dinh, Ingram, Inta, Isogai, Ivanov, Izumi, Jacobson, James, Jang,
  Jaranowski, Jesse, Johnson, Jones, Jones, Jonker, Ju, Kalmus, Kalogera,
  Kandhasamy, Kang, Kanner, Kasprzack, Kasturi, Katsavounidis, Katzman, Kaufer,
  Kaufman, Kawabe, Kawamura, Kawazoe, Keitel, Kelley, Kells, Keppel, Keresztes,
  Khalaidovski, Khalili, Khazanov, Kim, Kim, Kim, Kim, Kim, Kim, King, Kinzel,
  Kissel, Klimenko, Kline, Kokeyama, Kondrashov, Koranda, Korth, Kowalska,
  Kozak, Kringel, Krishnan, Kr\'olak, Kuehn, Kumar, Kumar, Kurdyumov, Kwee,
  Lam, Landry, Langley, Lantz, Lastzka, Lawrie, Lazzarini, Le~Roux, Leaci, Lee,
  Lee, Lee, Leong, Leonor, Leroy, Letendre, Lhuillier, Li, Li, Lindquist,
  Litvine, Liu, Liu, Lockerbie, Lodhia, Logue, Lorenzini, Loriette, Lormand,
  Losurdo, Lough, Lubinski, L\"uck, Lundgren, Macarthur, Macdonald,
  Machenschalk, MacInnis, Macleod, Mageswaran, Mailand, Majorana, Maksimovic,
  Malvezzi, Man, Mandel, Mandic, Mantovani, Marchesoni, Marion, M\'arka,
  M\'arka, Markosyan, Maros, Marque, Martelli, Martin, Martin, Marx, Mason,
  Masserot, Matichard, Matone, Matzner, Mavalvala, Mazzolo, McCarthy,
  McClelland, McGuire, McIntyre, McIver, Meadors, Mehmet, Meier, Melatos,
  Melissinos, Mendell, Men\'endez, Mercer, Meshkov, Messenger, Meyer, Miao,
  Michel, Milano, Miller, Minenkov, Mingarelli, Mitrofanov, Mitselmakher,
  Mittleman, Moe, Mohan, Mohapatra, Moraru, Moreno, Morgado, Morgia, Mori,
  Morriss, Mosca, Mossavi, Mours, Mow\char21{}Lowry, Mueller, Mueller,
  Mukherjee, Mullavey, M\"uller-Ebhardt, Munch, Murphy, Murray, Mytidis, Nash,
  Naticchioni, Necula, Nelson, Neri, Newton, Nguyen, Nishizawa, Nitz, Nocera,
  Nolting, Normandin, Nuttall, Ochsner, O'Dell, Oelker, Ogin, Oh, Oh,
  Oldenberg, O'Reilly, O'Shaughnessy, Osthelder, Ott, Ottaway, Ottens,
  Overmier, Owen, Page, Palladino, Palomba, Pan, Pankow, Paoletti, Paoletti,
  Papa, Parisi, Pasqualetti, Passaquieti, Passuello, Pedraza, Penn, Perreca,
  Persichetti, Phelps, Pichot, Pickenpack, Piergiovanni, Pierro, Pihlaja,
  Pinard, Pinto, Pitkin, Pletsch, Plissi, Poggiani, P\"old, Postiglione, Poux,
  Prato, Predoi, Prestegard, Price, Prijatelj, Principe, Privitera, Prodi,
  Prokhorov, Puncken, Punturo, Puppo, Quetschke, Quitzow-James, Raab, Rabeling,
  R\'acz, Radkins, Raffai, Rakhmanov, Ramet, Rankins, Rapagnani, Raymond, Re,
  Reed, Reed, Regimbau, Reid, Reitze, Ricci, Riesen, Riles, Roberts, Robertson,
  Robinet, Robinson, Robinson, Rocchi, Roddy, Rodriguez, Rodruck, Rolland,
  Rollins, Romano, Romie, Rosi\ifmmode~\acute{n}\else \'{n}\fi{}ska, R\"over,
  Rowan, R\"udiger, Ruggi, Ryan, Salemi, Sammut, Sandberg, Sankar, Sannibale,
  Santamar\'{i}a, Santiago-Prieto, Santostasi, Saracco, Sassolas,
  Sathyaprakash, Saulson, Savage, Schilling, Schnabel, Schofield, Schulz,
  Schutz, Schwinberg, Scott, Scott, Seifert, Sellers, Sentenac, Sergeev,
  Shaddock, Shaltev, Shapiro, Shawhan, Shoemaker, Sidery, Siemens, Sigg,
  Simakov, Singer, Singer, Sintes, Skelton, Slagmolen, Slutsky, Smith, Smith,
  Smith, Smith-Lefebvre, Somiya, Sorazu, Speirits, Sperandio, Stefszky,
  Steinert, Steinlechner, Steinlechner, Steplewski, Stochino, Stone, Strain,
  Strigin, Stroeer, Sturani, Stuver, Summerscales, Sung, Susmithan, Sutton,
  Swinkels, Szeifert, Tacca, Taffarello, Talukder, Tanner, Tarabrin, Taylor,
  ter Braack, Thomas, Thorne, Thorne, Thrane, Th\"uring, Titsler, Tokmakov,
  Tomlinson, Toncelli, Tonelli, Torre, Torres, Torrie, Tournefier, Travasso,
  Traylor, Tse, Ugolini, Vahlbruch, Vajente, van~den Brand, Van Den~Broeck,
  van~der Putten, van Veggel, Vass, Vasuth, Vaulin, Vavoulidis, Vecchio,
  Vedovato, Veitch, Veitch, Venkateswara, Verkindt, Vetrano, Vicer\'e, Villar,
  Vinet, Vitale, Vocca, Vorvick, Vyatchanin, Wade, Wade, Wade, Waldman,
  Wallace, Wan, Wang, Wang, Wanner, Ward, Was, Weinert, Weinstein, Weiss,
  Welborn, Wen, Wessels, West, Westphal, Wette, Whelan, Whitcomb, White,
  Whiting, Wiesner, Wilkinson, Willems, Williams, Williams, Willke, Wimmer,
  Winkelmann, Winkler, Wipf, Wiseman, Wittel, Woan, Wooley, Worden, Yablon,
  Yakushin, Yamamoto, Yamamoto, Yancey, Yang, Yeaton-Massey, Yoshida, Yvert,
  Zadro\ifmmode~\dot{z}\else \.{z}\fi{}ny, Zanolin, Zendri, Zhang, Zhang, Zhao,
  Zotov, Zucker, \& Zweizig}]{LIGO:2013}
Aasi, J., Abadie, J., Abbott, B.~P., {et~al.} 2013, Phys. Rev. D, 88, 062001

\bibitem[{{Abadie} {et~al.}(2010){Abadie}, {Abbott}, {Abbott}, {Abernathy},
  {Accadia}, \& et~al.}]{LIGOparamReview:2010}
{Abadie}, J., {Abbott}, B.~P., {Abbott}, R., {et~al.} 2010, Classical and
  Quantum Gravity, 27, 173001

\bibitem[{Acernese {et~al.}(2008)Acernese, Alshourbagy, Amico, Antonucci,
  Aoudia, Arun, Astone, Avino, Baggio, Ballardin, Barone, Barsotti, Barsuglia,
  Bauer, Bigotta, Birindelli, Bizouard, Boccara, Bondu, Bosi, Braccini,
  Bradaschia, Brillet, Brisson, Buskulic, Cagnoli, Calloni, Campagna,
  Carbognani, Cavalier, Cavalieri, Cella, Cesarini, Chassande-Mottin,
  Chatterji, Cleva, Coccia, Corda, Corsi, Cottone, Coulon, Cuoco, D'Antonio,
  Dari, Dattilo, Davier, Rosa, Prete, Fiore, Lieto, Emilio, Virgilio, Evans,
  Fafone, Ferrante, Fidecaro, Fiori, Flaminio, Fournier, Frasca, Frasconi,
  Gammaitoni, Garufi, Genin, Gennai, Giazotto, Granata, Greverie, Grosjean,
  Guidi, Hamdani, Hebri, Heitmann, Hello, Huet, Penna, Laval, Leroy, Letendre,
  Lopez, Lorenzini, Loriette, Losurdo, Mackowski, Majorana, Man, Mantovani,
  Marchesoni, Marion, Marque, Martelli, Masserot, Menzinger, Milano, Minenkov,
  Mohan, Moreau, Morgado, Mosca, Mours, Neri, Nocera, Pagliaroli, Palomba,
  Paoletti, Pardi, Pasqualetti, Passaquieti, Passuello, Piergiovanni, Pinard,
  Poggiani, Punturo, Puppo, Rabaste, Rapagnani, Regimbau, Remillieux, Ricci,
  Ricciardi, Rocchi, Rolland, Romano, Ruggi, Sentenac, Solimeno, Swinkels,
  Terenzi, Toncelli, Tonelli, Tournefier, Travasso, Vajente, van~den Brand,
  van~der Putten, Verkindt, Vetrano, Vicere, Vinet, Vocca, \&
  Yvert}]{Virgo:2008}
Acernese, F., Alshourbagy, M., Amico, P., {et~al.} 2008, Classical and Quantum
  Gravity, 25, 184001

\bibitem[{{Bhattacharya} {et~al.}(1992){Bhattacharya}, {Wijers}, {Hartman}, \&
  {Verbunt}}]{bwh+92}
{Bhattacharya}, D., {Wijers}, R.~A.~M.~J., {Hartman}, J.~W., \& {Verbunt}, F.
  1992, \aap, 254, 198

\bibitem[{Burke-Spolaor \& Bannister(2014)}]{BurkeSpolaorBannister:2014}
Burke-Spolaor, S., \& Bannister, K.~W. 2014, The Astrophysical Journal, 792, 19

\bibitem[{{Chu} {et~al.}(2015){Chu}, {Howell}, {Rowlinson}, {Gao}, {Zhang},
  {Tingay}, {Boer}, \& {Wen}}]{chu+2015}
{Chu}, Q., {Howell}, E.~J., {Rowlinson}, A., {et~al.} 2015, ArXiv e-prints,
  arXiv:1509.06876

\bibitem[{{Cordes} {et~al.}(2004){Cordes}, {Bhat}, {Hankins}, {McLaughlin}, \&
  {Kern}}]{cordes+04}
{Cordes}, J.~M., {Bhat}, N.~D.~R., {Hankins}, T.~H., {McLaughlin}, M.~A., \&
  {Kern}, J. 2004, \apj, 612, 375

\bibitem[{{Cordes} \& {Lazio}(2002)}]{CordesLazio:2002}
{Cordes}, J.~M., \& {Lazio}, T.~J.~W. 2002, ArXiv Astrophysics e-prints,
  astro-ph/0207156

\bibitem[{{Dominik} {et~al.}(2015){Dominik}, {Berti}, {O'Shaughnessy},
  {Mandel}, {Belczynski}, {Fryer}, {Holz}, {Bulik}, \&
  {Pannarale}}]{DominikEtAl:2014}
{Dominik}, M., {Berti}, E., {O'Shaughnessy}, R., {et~al.} 2015, \apj, 806, 263

\bibitem[{{Falcke} \& {Rezzolla}(2014)}]{FalckeRezzolla:2014}
{Falcke}, H., \& {Rezzolla}, L. 2014, \aap, 562, A137

\bibitem[{{Faucher-Gigu{\`e}re} \& {Kaspi}(2006)}]{fgk06}
{Faucher-Gigu{\`e}re}, C.-A., \& {Kaspi}, V.~M. 2006, \apj, 643, 332

\bibitem[{{Foucart} {et~al.}(2013){Foucart}, {Buchman}, {Duez}, {Grudich},
  {Kidder}, {MacDonald}, {Mroue}, {Pfeiffer}, {Scheel}, \& {Szilagyi}}]{fbd13}
{Foucart}, F., {Buchman}, L., {Duez}, M.~D., {et~al.} 2013, \prd, 88, 064017

\bibitem[{{Goldreich} \& {Julian}(1969)}]{gj69}
{Goldreich}, P., \& {Julian}, W.~H. 1969, \apj, 157, 869

\bibitem[{{Goldreich} \& {Lynden-Bell}(1969)}]{glb69}
{Goldreich}, P., \& {Lynden-Bell}, D. 1969, \apj, 156, 59

\bibitem[{{Gonthier} {et~al.}(2002){Gonthier}, {Ouellette}, {Berrier},
  {O'Brien}, \& {Harding}}]{gob+02}
{Gonthier}, P.~L., {Ouellette}, M.~S., {Berrier}, J., {O'Brien}, S., \&
  {Harding}, A.~K. 2002, \apj, 565, 482

\bibitem[{Gourgouliatos \& Cumming(2014)}]{GourgouliatosCumming:2014}
Gourgouliatos, K.~N., \& Cumming, A. 2014, Phys. Rev. Lett., 112, 171101

\bibitem[{{Hansen} \& {Lyutikov}(2001)}]{HansenLyutikov:2001}
{Hansen}, B.~M.~S., \& {Lyutikov}, M. 2001, \mnras, 322, 695

\bibitem[{Harry \& the LIGO Scientific~Collaboration(2010)}]{Harry:2010}
Harry, G.~M., \& the LIGO Scientific~Collaboration. 2010, Classical and Quantum
  Gravity, 27, 084006

\bibitem[{{Kasliwal} \& {Nissanke}(2014)}]{kasliwalNissanke:2014}
{Kasliwal}, M.~M., \& {Nissanke}, S. 2014, \apjl, 789, L5

\bibitem[{{Katz}(2014)}]{katz:2014}
{Katz}, J.~I. 2014, \prd, 89, 103009

\bibitem[{{Katz}(2015)}]{Katz:2015}
---. 2015, ArXiv e-prints, arXiv:1505.06220

\bibitem[{{Keane} {et~al.}(2012){Keane}, {Stappers}, {Kramer}, \&
  {Lyne}}]{KeaneEtAl:2012}
{Keane}, E.~F., {Stappers}, B.~W., {Kramer}, M., \& {Lyne}, A.~G. 2012, \mnras,
  425, L71

\bibitem[{{Lai}(2012)}]{Dong:2012}
{Lai}, D. 2012, \apjl, 757, L3

\bibitem[{{Li} \& {Paczy{\'n}ski}(1998)}]{LiPaczynski:1998}
{Li}, L.-X., \& {Paczy{\'n}ski}, B. 1998, \apjl, 507, L59

\bibitem[{{Lorimer} {et~al.}(1997){Lorimer}, {Bailes}, \& {Harrison}}]{lbh97}
{Lorimer}, D.~R., {Bailes}, M., \& {Harrison}, P.~A. 1997, \mnras, 289, 592

\bibitem[{{Lorimer} {et~al.}(2007){Lorimer}, {Bailes}, {McLaughlin},
  {Narkevic}, \& {Crawford}}]{LorimerEtAl:2007}
{Lorimer}, D.~R., {Bailes}, M., {McLaughlin}, M.~A., {Narkevic}, D.~J., \&
  {Crawford}, F. 2007, Science, 318, 777

\bibitem[{{Lorimer} \& {Kramer}(2012)}]{LorimerKramer:2012}
{Lorimer}, D.~R., \& {Kramer}, M. 2012, {Handbook of Pulsar Astronomy}
  (Cambridge University Press)

\bibitem[{{Luan} \& {Goldreich}(2014)}]{LuanGoldreich:2014}
{Luan}, J., \& {Goldreich}, P. 2014, \apjl, 785, L26

\bibitem[{{Lyutikov}(2011)}]{Lyutikov:2011}
{Lyutikov}, M. 2011, \prd, 83, 124035

\bibitem[{{McWilliams} \& {Levin}(2011)}]{McWilliamsLevin:2011}
{McWilliams}, S.~T., \& {Levin}, J. 2011, \apj, 742, 90

\bibitem[{{Metzger} \& {Berger}(2012)}]{MetzgerBerger:2012}
{Metzger}, B.~D., \& {Berger}, E. 2012, \apj, 746, 48

\bibitem[{{Nakar}(2007)}]{Nakar:2007}
{Nakar}, E. 2007, Physics Reports, 442, 166

\bibitem[{{Narayan} \& {Ostriker}(1990)}]{NarayanOstriker:1990}
{Narayan}, R., \& {Ostriker}, J.~P. 1990, \apj, 352, 222

\bibitem[{{Nissanke} {et~al.}(2013){Nissanke}, {Kasliwal}, \&
  {Georgieva}}]{nissanke+2013}
{Nissanke}, S., {Kasliwal}, M., \& {Georgieva}, A. 2013, \apj, 767, 124

\bibitem[{{Peters}(1964)}]{Peters1964}
{Peters}, P.~C. 1964, Physical Review, 136, 1224

\bibitem[{{Petroff} {et~al.}(2015){Petroff}, {Bailes}, {Barr}, {Barsdell},
  {Bhat}, {Bian}, {Burke-Spolaor}, {Caleb}, {Champion}, {Chandra}, {Da Costa},
  {Delvaux}, {Flynn}, {Gehrels}, {Greiner}, {Jameson}, {Johnston}, {Kasliwal},
  {Keane}, {Keller}, {Kocz}, {Kramer}, {Leloudas}, {Malesani}, {Mulchaey},
  {Ng}, {Ofek}, {Perley}, {Possenti}, {Schmidt}, {Shen}, {Stappers},
  {Tisserand}, {van Straten}, \& {Wolf}}]{PetroffEtAl:2015}
{Petroff}, E., {Bailes}, M., {Barr}, E.~D., {et~al.} 2015, \mnras, 447, 246

\bibitem[{{Popov} \& {Postnov}(2007)}]{pp07}
{Popov}, S.~B., \& {Postnov}, K.~A. 2007, ArXiv e-prints, arXiv:0710.2006

\bibitem[{{Popov} \& {Postnov}(2013)}]{PopovPostnov:2013}
---. 2013, ArXiv e-prints, arXiv:1307.4924

\bibitem[{{Ravi} {et~al.}(2015){Ravi}, {Shannon}, \& {Jameson}}]{RaviEtAl:2014}
{Ravi}, V., {Shannon}, R.~M., \& {Jameson}, A. 2015, \apjl, 799, L5

\bibitem[{{Sathyaprakash} {et~al.}(2012){Sathyaprakash}, {Abernathy},
  {Acernese}, {Ajith}, {Allen}, \& et~al.}]{ET:2012}
{Sathyaprakash}, B., {Abernathy}, M., {Acernese}, F., {et~al.} 2012, Classical
  and Quantum Gravity, 29, 124013

\bibitem[{Somiya(2012)}]{Somiya:2012}
Somiya, K. 2012, Classical and Quantum Gravity, 29, 124007

\bibitem[{{Spitler} {et~al.}(2014){Spitler}, {Cordes}, {Hessels}, {Lorimer},
  {McLaughlin}, {Chatterjee}, {Crawford}, {Deneva}, {Kaspi}, {Wharton},
  {Allen}, {Bogdanov}, {Brazier}, {Camilo}, {Freire}, {Jenet},
  {Karako-Argaman}, {Knispel}, {Lazarus}, {Lee}, {van Leeuwen}, {Lynch},
  {Ransom}, {Scholz}, {Siemens}, {Stairs}, {Stovall}, {Swiggum},
  {Venkataraman}, {Zhu}, {Aulbert}, \& {Fehrmann}}]{SpitlerEtAl:2014}
{Spitler}, L.~G., {Cordes}, J.~M., {Hessels}, J.~W.~T., {et~al.} 2014, \apj,
  790, 101

\bibitem[{{Stollman}(1987)}]{Stollman:1987}
{Stollman}, G.~M. 1987, \aap, 178, 143

\bibitem[{{Szary} {et~al.}(2014){Szary}, {Zhang}, {Melikidze}, {Gil}, \&
  {Xu}}]{szm+14}
{Szary}, A., {Zhang}, B., {Melikidze}, G.~I., {Gil}, J., \& {Xu}, R.-X. 2014,
  \apj, 784, 59

\bibitem[{{Thornton} {et~al.}(2013){Thornton}, {Stappers}, {Bailes},
  {Barsdell}, {Bates}, {Bhat}, {Burgay}, {Burke-Spolaor}, {Champion}, {Coster},
  {D'Amico}, {Jameson}, {Johnston}, {Keith}, {Kramer}, {Levin}, {Milia}, {Ng},
  {Possenti}, \& {van Straten}}]{ThorntonEtAl:2013}
{Thornton}, D., {Stappers}, B., {Bailes}, M., {et~al.} 2013, Science, 341, 53

\bibitem[{{Totani}(2013)}]{Totani:2013}
{Totani}, T. 2013, Publications of the Astronomical Society of Japan, 65, L12

\bibitem[{{Trott} {et~al.}(2013){Trott}, {Tingay}, \& {Wayth}}]{trott:2013}
{Trott}, C.~M., {Tingay}, S.~J., \& {Wayth}, R.~B. 2013, \apjl, 776, L16

\bibitem[{{Tsang} {et~al.}(2012){Tsang}, {Read}, {Hinderer}, {Piro}, \&
  {Bondarescu}}]{tsang+2012}
{Tsang}, D., {Read}, J.~S., {Hinderer}, T., {Piro}, A.~L., \& {Bondarescu}, R.
  2012, Physical Review Letters, 108, 011102

\bibitem[{{Veitch} {et~al.}(2015){Veitch}, {Raymond}, {Farr}, {Farr}, {Graff},
  {Vitale}, {Aylott}, {Blackburn}, {Christensen}, {Coughlin}, {Del Pozzo},
  {Feroz}, {Gair}, {Haster}, {Kalogera}, {Littenberg}, {Mandel},
  {O'Shaughnessy}, {Pitkin}, {Rodriguez}, {R{\"o}ver}, {Sidery}, {Smith}, {Van
  Der Sluys}, {Vecchio}, {Vousden}, \& {Wade}}]{vrf+15}
{Veitch}, J., {Raymond}, V., {Farr}, B., {et~al.} 2015, \prd, 91, 042003

\bibitem[{{Yancey} {et~al.}(2015){Yancey}, {Bear}, {Akukwe}, {Chen}, {Dowell},
  {Gough}, {Kanner}, {Kavic}, {Obenberger}, {Shawhan}, {Simonetti}, \& {-Wei
  Tsai}}]{yba+15}
{Yancey}, C.~C., {Bear}, B.~E., {Akukwe}, B., {et~al.} 2015, \apj, 812, 168

\end{thebibliography}

\end{document}